\documentclass[letterpaper]{article}
\usepackage{natbib,alife13}
\usepackage{multirow}
\usepackage{siunitx}
\usepackage{caption}
\usepackage{fancyhdr}

\title{Behavioral Strategy Chases Promote the Evolution of Prey Intelligence}

\author{Aaron P Wagner$^{1}$, Luis Zaman$^{1,2,3}$, Ian Dworkin$^{1,3,4}$, Charles Ofria$^{1,2,3}$ \\
\mbox{}\\
$^{1}$ BEACON Center for the Study of Evolution in Action, Michigan State University, East Lansing, MI 48824, USA\\
$^{2}$ Department of Computer Science and Engineering, Michigan State University, East Lansing, MI 48824, USA \\
$^{3}$ Program in Ecology, Evolutionary Biology, and Behavior, Michigan State University, East Lansing, MI 48824, USA\\
$^{4}$ Department of Zoology, Michigan State University, East Lansing, MI 48824, USA\\
}

\begin{document}

\maketitle

\fancypagestyle{empty}{%
  \fancyhf{}% Clear header/footer
  \fancyhead[L]{\textit{ }}% Your journal/note
  %\fancyhead[R]{arXiv:1310.1369}% Your logo/image
}

{\section{Abstract}
\parindent0pt \textbf{Predator-prey coevolution is commonly thought to result in reciprocal arms races that produce increasingly extreme and complex traits. However, such directional change is not inevitable. Here, we provide evidence for a previously undemonstrated dynamic that we call “strategy chases,” wherein populations explore strategies with similar levels of complexity, but differing behaviorally. Indeed, in populations of evolving digital organisms, as prey evolved more effective predator-avoidance strategies, they explored a wider range of behavioral strategies in addition to exhibiting increased levels of behavioral complexity. Furthermore, coevolved prey became more adept in foraging, evidently through coopting components of explored sense-and-flee avoidance strategies into sense-and-retrieve foraging strategies. Specifically, we demonstrate that coevolution induced non-escalating exploration of behavioral space, corresponding with significant evolutionary advancements, including increasingly intelligent behavioral strategies.}}
\\
\\
\section{Introduction}
Dawkins and Krebs (1) famously proposed that Red Queen Dynamics (2) in antagonistic systems should produce reciprocal evolutionary arms races (3). This hypothesis predicts that the interacting species coevolve traits in a tit-for-tat exchange of increasingly extreme adaptations and counter-adaptations. I.e. “swords get sharper, so shields get thicker, so swords get sharper still”. While versions of this arms race model often dominate popular explanations and scientific expectations (4), there is support for alternative and non-escalating coevolutionary mechanisms, including trait cycling (5) and defense-preference alternation (6,7). However, the potential importance of non-escalating coevolutionary exploration of behavioral strategies remains largely unconsidered and untested.

In order to understand how expectations for behavioral phenotypes may differ from other traits, consider the common expectations under arms race models. Arms races are typically couched in terms of effects on the complexity (8,9) of an individual aspect of morphology or behavior (1,4,10-14). E.g., stronger claws vs. thicker shells or speed of chase vs. speed of flight. In such a model, directional selection is predicted to drive increased complexity in both players over evolutionary time. To test for this, traits can be evaluated in terms of how much information they incorporate about the environment (e.g., shell thickness reflecting predator capabilities for crushing; after 15). Such a directional model requires that potential evolutionary responses fall along few axes, and the antagonistic nature of predator-prey interactions typically ensures that only one direction of travel along an axis is viable. For example, all else equal, just forming thinner shells is not a viable evolutionary response to increased predator crushing strength. However, phenotypes of equivalent behavioral complexity, as measured by the use of information from the environment to inform actions, can carry different fitness effects: a grey moth with an expressed behavioral preference for perching on grey trees is likely safer, but no more behaviorally complex, than a grey moth expressing a preference for perching on black trees. Additionally, behavior is not defined by single, isolated actions, but a series of interrelated actions. Given the numbers and combinations of potential actions, the dimensionality of options for even simple behavioral strategies can be vast. For example, while the complexity of prey responses to coursing predators could increase over evolutionary time, viable alternative flight behaviors could also consist of zig-zagging, hiding, or sudden stops and redirections, as well as variations of each. If each strategy incorporates the same number of informed actions, they are of equal behavioral complexity.

Since equally complex strategies are unlikely to be uniformly effective against a given predator, we should expect evolution to produce exploratory ‘chases’ through behavioral options space as often as producing arms races for increasing complexity. While a number of studies (2-7,11-14,16-19) have discussed escalating arms races and non-escalating alternatives in antagonistic interactions, we are not aware of any that have examined the relative importance of evolutionary behavioral strategy exploration in defining the outcomes of predator-prey coevolution. A major constraint on testing for these processes is the inherent difficulty of the simultaneous, detailed, and prolonged experimental study of behavior in predators and prey (20), particularly over evolutionary time. However, computational systems permit this sort of inquiry. Specifically, the experimental evolution software Avida (21) carries all of the benefits of evolutionary simulations (e.g., rapid generation times and full control over experimental environments), without incorporating explicit fitness functions to artificially select individuals for reproduction. Importantly, Avida does not merely simulate evolution (22), nor does it carry the assumptions inherent to selection regimes and other mathematical models (21): a digital organism in Avida has a genome subject to random mutations that are inherited by its offspring, as well as a fitness determined by realized competitive abilities to survive, collect needed resources, and produce offspring. Uniquely among computational systems, this combination allows for unrestricted, unsupervised, and non-deterministic evolution via natural selection, and direct testing of biological hypotheses (21). Here we use the Avida system to show that coevolution among predators and prey produces both escalating arms races and non-escalating chases through behavioral options.

Avida populations exhibit a rich range of evolutionary dynamics and have been used to understand many factors behind the evolution of complexity (15,23), including its emergence as a consequence of antagonistic host-parasite interactions (24). The genomes of the digital organisms consist of low-level computational instructions, including those for environmental sensing, controlling the order and conditions of instruction execution, and for reproduction (at the cost of consumed resources). During reproduction, mutations can occur, producing genetic differences between parent and offspring genotypes. We modified Avida to include a predation instruction (25) that, if mutated into a genome, makes the carrier capable of predating other organisms. If an organism executes that instruction and makes a kill, the attacking organism is classified as a predator. All organisms were required to consume enough resources to meet a threshold for reproduction. Accordingly, prey needed to locate and consume food in the environment, while predators needed to locate, successfully attack, and consume multiple prey. Importantly, predators are simply organisms that evolved to eat other organisms, sharing a common genetic instruction set with prey and interacting in the same ways with their environment. As in nature, it is only evolved changes in genetic sequences and behaviors that differentiate predators from their prey (see Fig. S1 and Movie S1). 

\makeatletter
\setlength{\@fptop}{5pt}
\begin{figure}[!b]
\centering

\includegraphics[width=2.9in]{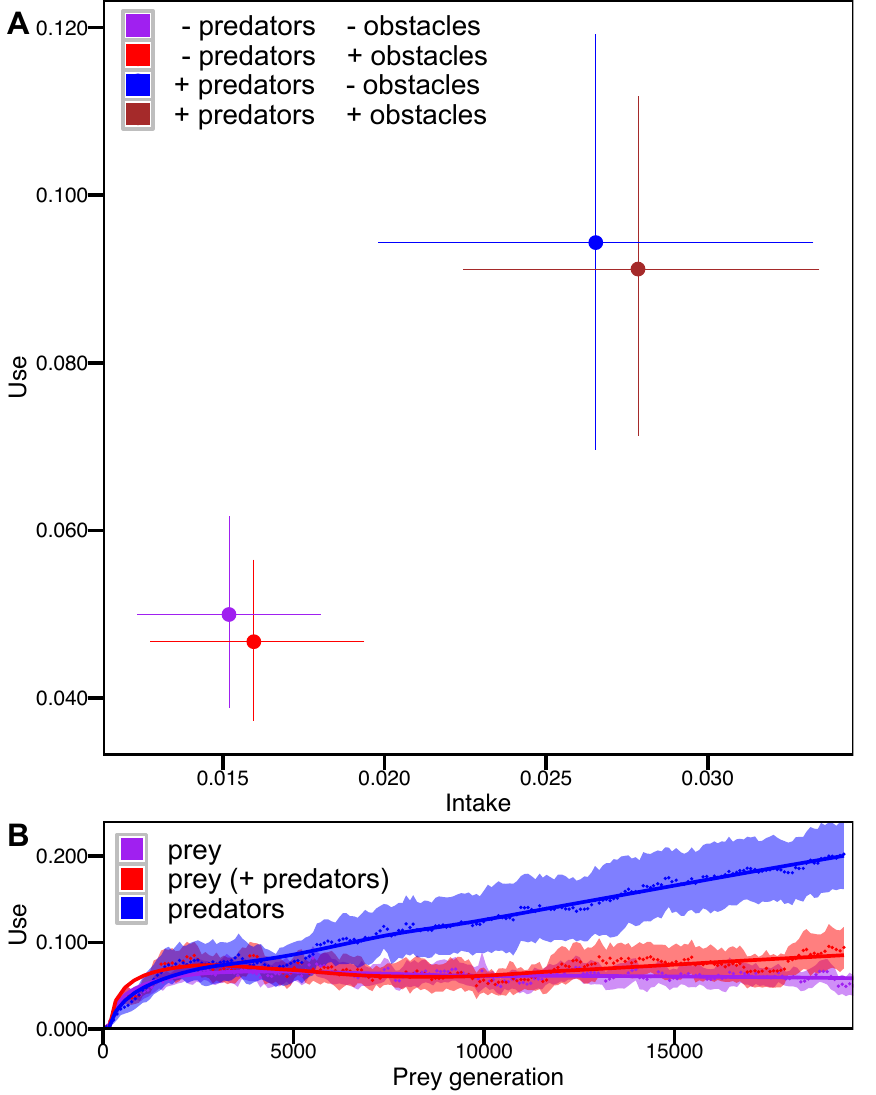}
\caption{ \textbf{Coevolution promotes increased prey behavioral intelligence and complexity. (A)} Coevolution with predators significantly increases both the rate of sensory intake and the rate of information use (realized behavioral intelligence), while abiotic environmental complexity (obstacles) has little effect. Data shown are from the final evolutionary time-point. \textbf{(B)} While predator behavioral intelligence increases linearly over the course of evolution, coevolving prey evolve to use sensory information later and at a lower rate. Lines are LOESS fits. Mean prey generation times were 102.21 updates ($\pm$ 0.11 se), with predator-to-prey generation time ratios of 2.49:1 ($\pm$ 0.06 se). Shaded regions and error bars are bootstrapped 95\% confidence intervals.
}
\label{fig1}
\end{figure}

We initialized all evolutionary trials with prey that randomly moved about the environment, indiscriminately attempting to consume resources and reproduce. Among potential adaptive targets, evolution could refine these simple behaviors via adaptations for sensing and responding to objects (i.e. food, organisms, barriers) and more controlled navigation or avoidance strategies. We performed evolutionary trials conducted with (\textit{Pred+}) and without (\textit{Pred-}) the possibility of predator coevolution, and monitored both frequency of sensor use (Fig. 1) and behavioral intelligence and complexity, defined as the proportion of genetic actions (decisions) that relied on sensory information.

\section{Results and Discussion}
After two million evolutionary time steps ($\approx$19,500 prey generations), termed updates, sensor use was higher for prey populations evolved with predators (mean=0.027, 95\% CI: 0.019,0.033) vs. those evolved in the absence of predators (mean=0.015, 95\% CI: 0.012,0.018; Kruskal-Wallis \textit{p}=0.033). Likewise, behavioral intelligence and complexity, was also higher in prey populations evolved with predators (\textit{Pred+}: mean=0.094, 95\% CI: 0.070,0.120; \textit{Pred-}: mean=0.050, 95\% CI: 0.038,0.061; Kruskal-Wallis \textit{p}=0.005). In contrast, behavioral intelligence and complexity did not change in response to more complex abiotic environments: distributing barriers (Fig. S2) throughout the environment had no detectable effect on evolved levels of behavioral complexity (Fig. S3). 

\makeatletter
\setlength{\@fptop}{5pt}
\begin{figure}[!b]
\centering

\includegraphics[width=2.9in]{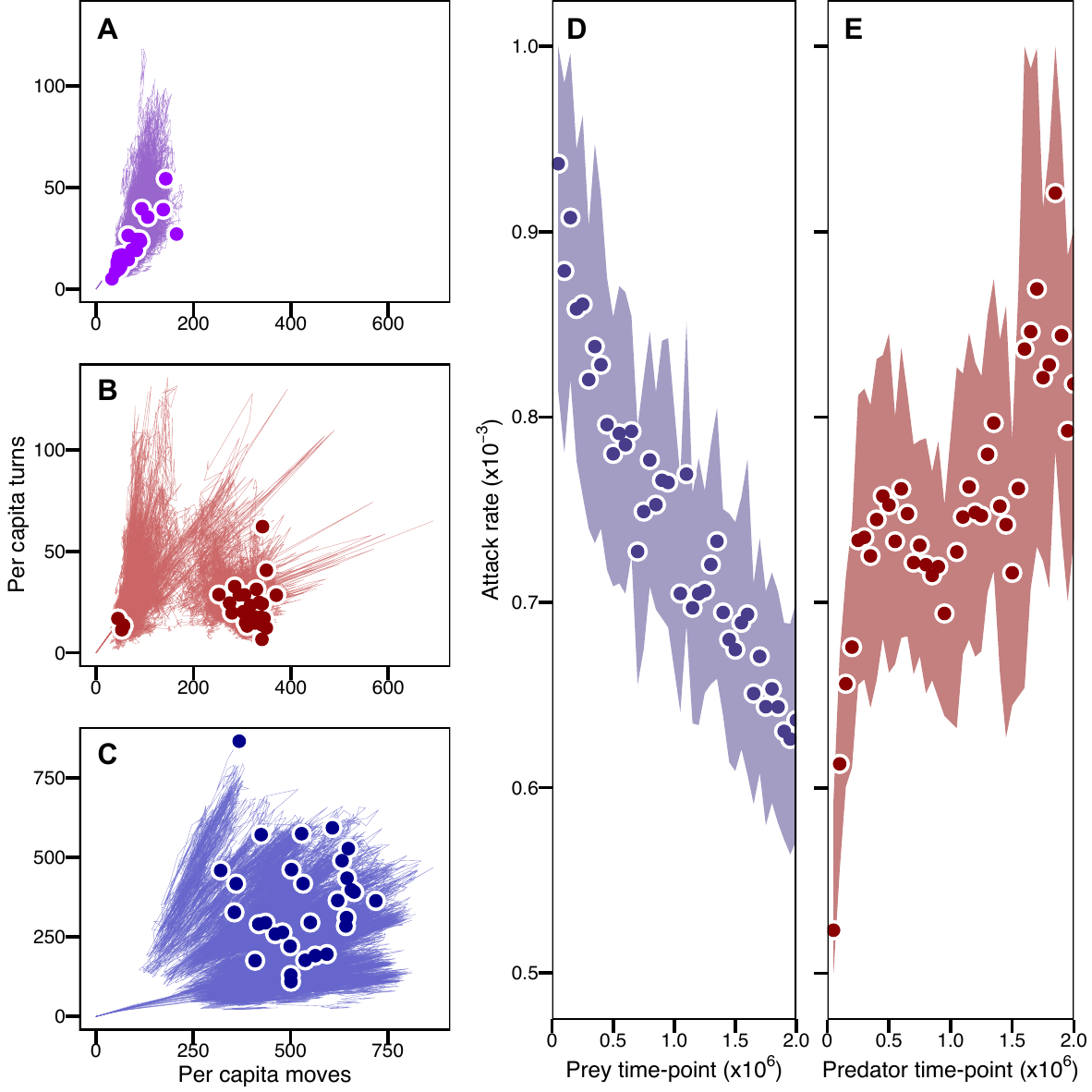}
\caption{ \textbf{Coevolving populations explore more behavioral strategies while improving performance.} Shown are mean number of turns and moves taken in each population of \textbf{(A)} \textit{Pred-} prey, \textbf{(B)} \textit{Pred+} prey, and \textbf{(C)} predators (note change in scale) over evolutionary time. Points denote final behaviors in each of 30 trials. Even when returning to the low-movement and low-turn behavioral strategies nearer to that of the naïve ancestor (at the origin), \textit{Pred+} prey populations explore parts of behavioral space never investigated by \textit{Pred-} prey. For all but three \textit{Pred+} prey populations, that exploration leads to a major transition, allowing them to into an area defined by high movement rates. \textbf{(D)} Mean attack rates when prey from different time-points are reintroduced with predators from the middle of the same evolutionary timeline are highest when predators face the most naive prey and lowest when facing the most fully evolved prey. \textbf{(E)} Likewise, attack rates increase when predators from each time-point face prey from the middle of their evolutionary timeline \textbf{(E)}. X-axes in \textbf{(D)} and \textbf{(E)} indicate the time-point (update) from which the indicated populations were drawn; shaded regions are bootstrapped 95\% confidence intervals.
}
\label{fig2}
\end{figure}

As predicted, prey coevolving with predators explored a greater area of behavioral space, as described by executed rates of moves and turns (the only two possible ‘physical’ behavioral actions for prey): \textit{Pred+} prey populations made frequent forays into new areas of move-turn behavioral space, while \textit{Pred-} prey remained in a much smaller sub-area (Fig. 2, see also Fig. S4). As a consequence of this behavioral exploration, 27 of the 30 prey populations coevolving with predators discovered, moved to, and then remained in, an area of behavioral space clearly separated from that used by naïve populations (i.e. \textit{Pred-} populations and evolutionarily young \textit{Pred+} populations). As a measure of their true and realized extent of exploration, cumulative lengths of the paths connecting observation points in this move-turn behavioral space were substantially longer for prey coevolving with predators than in counterpart populations (\textit{Pred+}: median path length=11,222 steps, range: 6026-26,323; \textit{Pred-}: median=6,399, range: 3,387-9,079), and even longer for predators themselves (median=74,243, range: 49,270-92,223) (Fig. S5).

In addition to exploring more of behavioral space and taking in and using more information in making decisions, prey coevolving with predators exhibited increasingly effective predator avoidance strategies (Fig. 2 and Movie S1): attack rates decreased for predators that were reintroduced into time-shift trials [26,27] with the prey from earlier vs. later in their evolutionary history. Likewise, hunting performance of predators clearly improved over time, as measured by presenting predator populations along each evolutionary timeline with the prey from the middle of that timeline. Additionally, attack rates on evolving prey declined at a constant rate (mean=0.937, 95\% CI: 0.818,1.057, at the first sample, declining to mean=0.636, 95\% CI: 0.571,0.702, at the final sample), even while use of sensory information exhibited minimal change (e.g., the second quarter of the evolutionary timelines, Fig. 1), indicating that prey continued to explore new and more effective anti-predator behavioral strategies even in the absence of increased behavioral intelligence and complexity. Similarly, there was no indication of a movement arms race: while \textit{Pred+} prey settled in an area of behavioral space defined by relatively high rates of movement, final movement rates for coevolving species were below explored maxima (Fig. 2, Fig. S4). Furthermore, in behavioral assays, removal of predators resulted in similar declines in prey movement (a proxy for length of flight responses) over  most of evolutionary time (mean=15.453\%, 95\% CI: -45.030,15.180, movement decline with predators removed at update 50,000 vs. mean=21.120\%, 95\% CI: -41.322, -2.367, decline if removed at the final update; Fig. S6).

\makeatletter
\setlength{\@fptop}{5pt}
\begin{figure}[!t]
\centering

\includegraphics[width=2.9in]{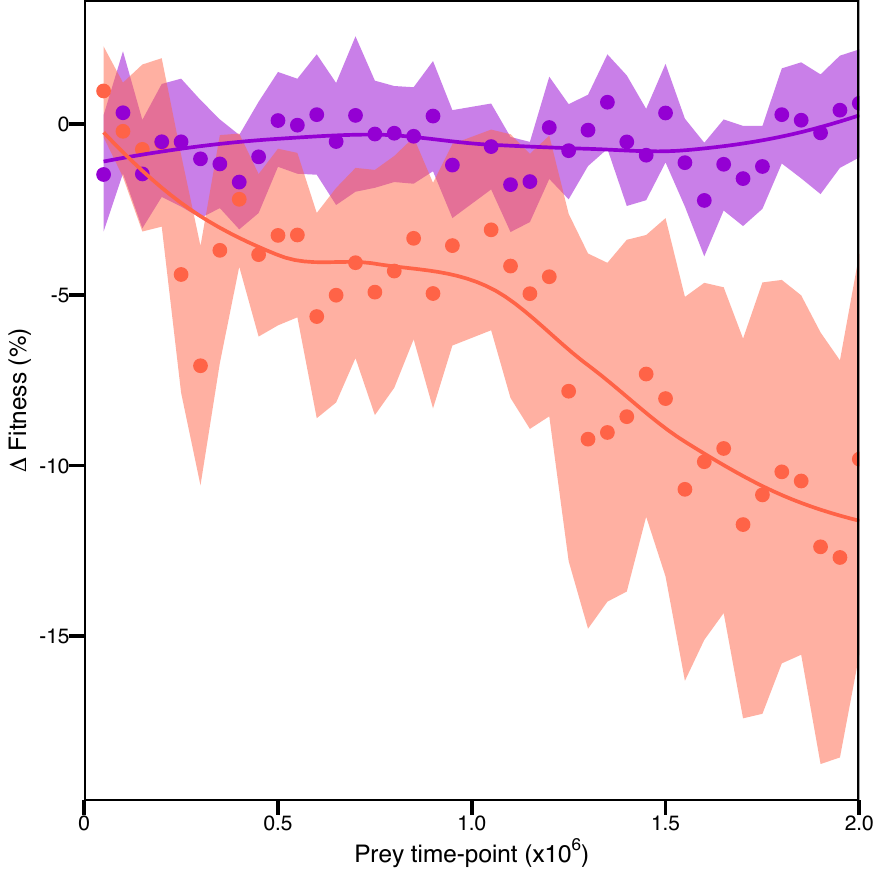}
\caption{ \textbf{Coevolution promotes intelligent use of sensory data in foraging decisions.} Shown are mean per-population changes in fitness (reflecting foraging success and gestation time) for prey before and after being blinded to resources. Prey evolved without predators (top, purple) exhibit limited declines in foraging success, indicating a lack of reliance on sensor information. In contrast, prey coevolved with predators (red) experience significant and increasing declines in success over evolutionary time. Shaded regions are bootstrapped 95\% confidence intervals. Lines are LOESS fits.
}
\label{fig3}
\end{figure}

We hypothesized that the evolution of behaviorally intelligent traits improving predator avoidance would also result in increased use of sensory data by prey for foraging. Indeed, prey coevolved with predators demonstrated a substantial reliance on information about their environment in making foraging decisions, and increasingly so over evolutionary time (Fig. 3): in additional behavioral assays, the “blinding” of prey to food resources resulted in a mean fitness (the quotient of lifetime food intake by replication time) decline of 0.968\% (95\% CI: -0.380,2.277) for populations tested at update 50,000, and a decline of 9.812\% (95\% CI: -15.926,-4.058) for fully evolved populations. In contrast, the blinding of prey evolved in the absence of predators decreased their fitness only slightly, and with little change in the magnitude of that effect over time (initial mean=-1.478\%, 95\% CI: -3.212,0.200, decline vs. 0.608, 95\% CI: -1.000,2.248, decline at the final update). Hence greater evolved use of information about the environment contributed significantly to prey fitness, beyond its importance for predator avoidance. 

Finally, the three coevolutionary effects on prey (increased information intake, use of information in decision making, and broader behavioral strategy exploration) also increased prey competitiveness. Specifically, we competed all \textit{Pred+} prey populations against all \textit{Pred-} populations in new, predator-free environments. At the end of competition, the descendants of prey coevolved with predators represented the majority in most populations (Fig. 4; \textit{Pred+}: 23.5 median in-majority counts, 95\% CI: 21.487-25.953; \textit{Pred-}: median 7, 95\% CI: 3.827-10.440; Kruskal-Wallis p$<$0.001). The competitive performance of prey coevolved with predators was further pronounced in additional trials with a 75\% reduction in resource regrowth rates. Thus, prey evolved with predators proved to be more adept and competitive in foraging than prey evolved without predators, including in the very environments one would otherwise expect the latter, not the former, to be more closely adapted. This appears to be a consequence of a reciprocal evolutionary relationship in which, as prey become better at sensing and reacting to predators, they more readily evolve to become better at sensing and reacting to resources, which further increases evolutionary discovery of adaptations for responsiveness to predators (Fig. S7).  

\makeatletter
\setlength{\@fptop}{5pt}
\begin{figure}[!t]
\centering

\includegraphics[width=2.9in]{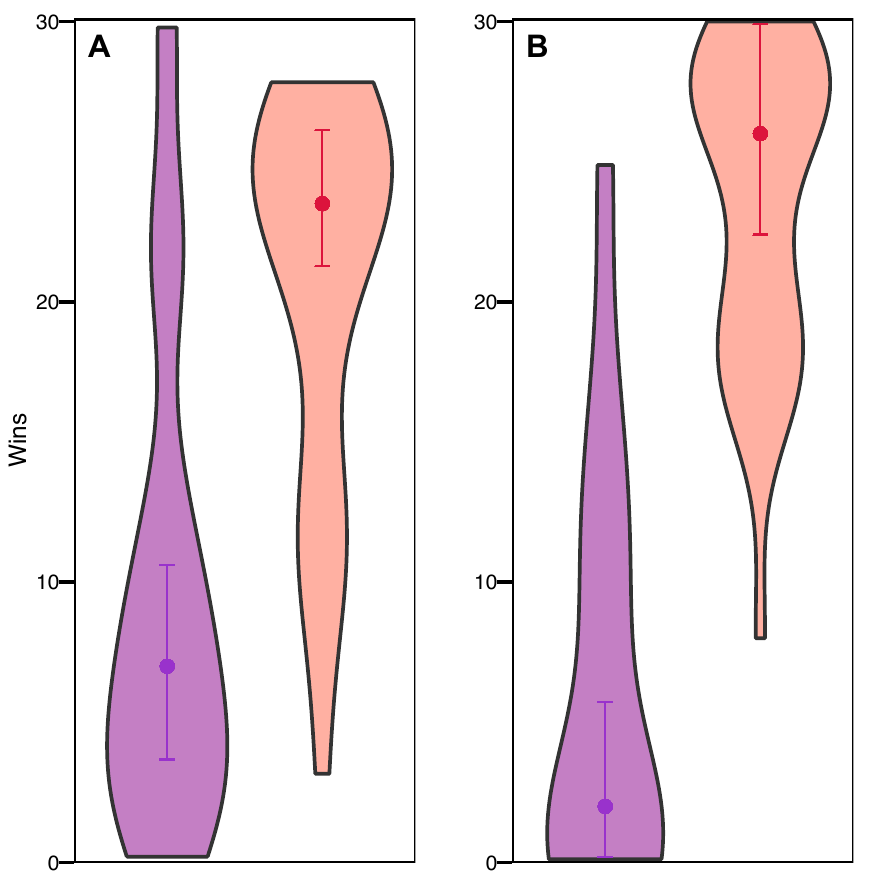}
\caption{ \textbf{Coevolution enhances prey competitiveness and behavioral flexibility. (A)} Results of competition between prey evolved with and without predators; shown are numbers of competitions in which \textit{Pred-} prey (purple) and \textit{Pred+} prey (red) dominated at the end of competition (30 competitions per population). \textbf{(B)} Results of competition in a more extreme environment in which resource regrowth rates were 25\% of that used in the evolutionary trials. In both environments, prey coevolved with predators dominated most competitions, with their competitiveness enhanced in the novel reduced growth environment. Points indicate medians. Error bars are bootstrapped 95\% confidence intervals. Shaded areas show the full distribution of per-population per-treatment wins. 
}
\label{fig4}
\end{figure}

Coevolution with predators produced more behaviorally complex and behaviorally intelligent prey. However, prey performance continued to improve even when complexity indicators did not. Instead, we observed an ongoing exploration of equally complex behaviors. Unlike pure arms races, such exploration of behavioral options need not be directional, nor is it as directly and tightly constrained as are physical traits (e.g., as in 17,28). While the extent of reciprocity in this process remains unexamined (2,4,29,30), we have demonstrated that such chases do produce significant evolutionary advancements, including early forms of behavioral intelligence producing more fit and competitive populations. We expect additional examination of the interplay between ecological interactions (31-33) and the exploration of behavioral strategy spaces will further highlight its importance in the evolutionary discovery of key innovations. 

\section{Methods}
\subsection{Environment}
We initialized each evolutionary trial with nine simple, identical prey organisms that randomly moved about the environment, blindly attempting to collect resources and reproduce. Each of the 30 evolutionary trials (of each treatment) was conducted for 2 million updates ($\approx$19,500 prey generations). Updates are the unit of time in Avida and one update is defined as the time required for each organism, on average, to execute 30 instructions.

All experiments were conducted in bounded grid-worlds of 251 by 251 cells. Each cell could contain up to one unit of food. When a prey fed from a cell, the prey consumed that full unit and the resource would then regenerate at a rate of 0.01 units per update. Thus any one cell could be fed from no more than once per 100 updates. Organisms were required to consume ten units of resources from the environment before they could reproduce. 

In the treatments that included barriers (which block movement), we created 25 pairs of barriers (Fig. S2). Within each pair, one barrier extended north to south, and the other barrier extended east to west, intersecting at their north and west ends, respectively. Intersection points were separated by 50 cells on each axis, leaving 30 cells between the end of a barrier in one pair and the intersection point of the next pair. Pairs were placed in five columns and five rows, with the northern-most row along the northern boundary of the grid (so that, for this row, the north-south barrier extended downward from the northern boundary, while the east-west barrier lay along the boundary) and the western-most column along the western boundary.

\subsection{Reproduction}
Provided that the organism had consumed sufficient resources and was old enough (minimum = 100 updates), reproduction occurred when an organism executed a single reproduction instruction, i.e. organisms used a composite instruction and were not required to copy individual instructions as in traditional configurations of Avida. For each new offspring genome, there was a 25\% chance of a single substitution mutation occurring, and 5\% chances for single insertion and deletion mutations occurring, independently. Genome lengths were unrestricted. New genetic mutations were suppressed in all reintroduction and competition experiments. Whereas most instructions took 1/30th of an update to execute (= 1 ‘cycle’), reproduction required a full update to complete.

New organisms were born into the cell faced by their parent. To limit population-size artifacts, populations evolving in the absence of predators were limited to 700 organisms. When a new birth would have caused the population level to exceed this cap, a random organism (other than the parent) was removed from the existing population. Organisms older than 500 updates were also removed.

\subsection{Predation}
Predators attack prey via the execution of a single instruction. If there is a prey in front of the predator and a kill is made, predators consume the prey with a conversion efficiency of 10\%. E.g., a predator would gain one unit of resource from consuming a prey that had eaten ten units of environmental resources. Reproduction for both predators and prey is limited by resource consumption: the faster an organism gathers food, the sooner it will be able to reproduce. For predators, mutations allowing for more effective location, pursuit, and capture of prey will therefor provide evolutionary advantages. Likewise, mutations in prey that improve foraging efficiency or predator avoidance provide selective advantages. In previous work in Avida, each cell in the world could hold only one organism. Here, however, there is no limit on the number of organisms per cell. Consequently, populations are limited from the bottom-up by resources, by setting explicit population caps, or, for prey, by top-down predation pressure. Because the experimental systems were effectively closed, and in order to allow for consistency in prey densities across trials and treatments, we prevented predator attacks from succeeding if the prey population fell below a minimum threshold of 700. In practice, this means repeated attacks could be required to make a kill. For each failed attack, the targeted prey was ‘injured’ via a 10\% reduction of their consumed, stored resource. 

\subsection{Forager Types}
All organisms were born in a neutral ‘juvenile’ state. Organisms could then alter their classification to become a predator or prey by executing a specific classifier instruction with the appropriate predator or prey value identifier in the modifying register. Organisms had to be classified as prey to consume environmental resources. Organisms were always classified as predators as soon as they had successfully executed any attack on a prey. Alternatively, organisms could adopt the forager classification of their parent if the parent had executed a ‘teach’ instruction and the (offspring) organism executed a ‘learn’ instruction. While prey could become predatory, predators were prevented from being reclassified as prey during their lifetime. In practice, success in the former was rare once predator and prey behaviors diverged significantly (which occurred very early in the evolutionary timelines, see Fig. 2 and Fig. S1) and each became more efficient in its own niche.  

\subsection{Sensors}
Organisms could evolve the use and control of environmental sensors capable of providing information about objects. Each organism’s area of vision was limited to its front octant out to a distance of 10 cells. Objects in the environment included other organisms, food and, in specified treatments, barriers blocking movement. Walls were also placed around the outer perimeter of the grid world, making the boundary detectable by organisms.

The capacities of the sensors were designed to allow organisms the ability to evolve extensive capacities for sight. Dependent on their evolved behavior, organisms could set and use integer values in four of their internal registers to query sensors for information, specifying: 1) the type of object they were looking for, including predators vs. prey, 2) the maximum distance to look to, 3) whether they were looking for the closest object of that type, or a count of all objects of that type in their visual field, 4) any specific instance of the type of object sought (e.g., a particular known organism). Eight integer outputs were returned by every use of a sensor: 1) the type of object searched for, 2) the distance to the object, or distance used if nothing was visible, 3) whether the closest object or a count of objects was sought, 4) the specific instance of an object that was sought, if specified in the input controls, 5) the count of objects of the correct type seen, 6) the values of the objects seen, 7) the identity of the object seen, 8) in a search for organisms, the type of organism seen (predator vs. prey). In essence, the sensors could become perfect eyes. However, they are useless (and potentially detrimental) unless organisms evolve mechanisms for controlling what information is processed from visual inputs. A complete list of sensor default behaviors is available in the Avida documentation.

\subsection{Hardware}
In Avida, the virtual hardware defines critical aspects of an organism’s construction, e.g., memory registers, potential instructions, and genome execution rules. We used the EX hardware [34], modified to include the eight registers needed for organisms to control their sensors and to allow up to four parallel execution threads. Threads were created if an organism executed a ‘fork’ instruction. Any instruction occurring in the genome between the fork and an ‘end-thread’ instruction were effectively copied to a second genome execution stream. Each thread also maintained its own complement of registers and a single stack (there was also one stack common to all threads). Each cycle, the current instruction for each thread was executed, in the order that the threads were created. Additional instructions were also available for threads to pause their own execution until certain values appeared in the registers of other threads. Beyond new instructions for predation and thread control, the instruction set also included instructions for detecting an organism’s heading (i.e. a compass), rotating multiple times, and rotating until a specific organism (detected via sensors and remembered) came back into view. 

\subsection{Complexity and Intelligence}
We measured potential complexity as the mean proportion of per-capita, total lifetime instructions executed that were sensing instruction executions, i.e. the level of information intake [8,9]. We measured realized behavioral complexity and behavioral intelligence as the mean proportion of instructions that used data originating from the sensors as regulatory or modifying inputs, i.e. the extent to which information was used and incorporated into decisions and actions [35].

\subsection{Behavioral Exploration}
We measured behavioral exploration in an x-y plane of per-capita moves and turns, recorded every 1000 updates for every population. The travel distance between recorded points was calculated (using the Pythagorean theorem) as the square root of the sum of the squared difference in per-capita turns and the squared difference in per-capita moves. Total explored distance, or path length, for each population was the sum of these distances over the two million updates of evolution (sum of 2,000 distances per population).

\subsection{Time-shifts}
We saved complete records of the genomes and birth locations of all living organisms every 50,000 updates during each evolutionary trial. To limit any potential artifacts related to location within the grid-world or age and developmental state, for all reintroductions, organisms were placed at their original birth location and with all internal states (e.g., memory) reset, as it is in new births, but retaining information about the organisms’ parents (e.g., whether the parent had executed a ‘teach’ instruction). 

To evaluate changes in prey abilities to avoid predators, each predator population from one million updates was reintroduced, in turn, along with the prey from each time point of the same trial. Likewise, to evaluate changes in predator abilities to catch prey, we reintroduced each saved predator population with the prey population from the middle of their evolutionary timeline. We then measured attack rates as the proportion of all lifetime instructions that were successful attacks for the parents of the predators alive at 1000 updates post-reintroduction (data from parents are used to allow evaluation over complete lifetimes). 

\subsection{Foraging Decisions}
We reintroduced each saved prey population, evolved with and without predators, into predator-free environments and measured mean fitness at 1000 updates. Fitness in Avida is calculated as lifetime food intake divided by gestation time (in updates). We then altered the sensors so that they would always return signals indicating the equivalent of ‘no food seen’ in response to an organism’s attempts to look for food, and again evaluated fitness in new reintroduction trials of the same source prey populations. Because the only variable changed across these two assays was the ability to acquire and respond to visual information about resources, we used the per-population changes in fitness as our measure of the importance of that knowledge in informing foraging decisions.

\subsection{Foraging Competitions}
To compete prey coevolved with predators against prey evolved without predators, each of the final prey populations from the evolutionary trials was paired once with each of the 30 final prey populations from the opposing treatment and reintroduced into a new environment. For each population, we then counted the number of competitions in which its descendants constituted the majority of the final total composite population after 200 generations of competition in environments with 100\% and 25\% of the resource regrowth rates used in the evolutionary trials. 

\subsection{Software}
We used Avida version 2.12 for all experiments. Data were post-processed using Python 2.7.1. Statistical analyses and plotting were conducted in R version 2.15.2 using the ggplot2 and boot libraries.

\section{References}
\hspace*{7pt} 1. Dawkins, R. $\&$ Krebs, J.R. Arms races between and within Species. Proc. R. Soc. Lond. B 205, 489-511 (1979)

2. van Valen, L. A new evolutionary law. Evol. Theory 1, 1-30 (1973)

3. Cott, H.B. Adaptive Coloration in Animals. London:Methuen (1940)

4. Abrams, P.A. Is predator-prey coevolution an arms race? TREE 1, 108-110 (1986)

5. Dieckmann, U., Marrow, P. $\&$ Law, R. J. Theor. Biol. 176, 91-102 (1995).

6. Davies, N.B. $\&$ Brooke, M.D. J. Animal Ecol. 58, 207-224 (1989).

7. Nuismer, S.L. $\&$ Thompson, J.N. Evolution 60, 2207-2217 (2006).

8. Adami C., Ofria, C. $\&$ Collier, T.C.. Evolution of biological complexity. PNAS 97, 4463-4468 (2000)

9. Adami, C. What is complexity? BioEssays 24, 1085-1094 (2002)

10. Vermeij, J. The evolutionary interaction among species: selection, escalation, and coevolution. Ann. Rev. Ecol. Syst. 25, 219-236 (1994)

11. Brodie, E.D. III $\&$ Brodie, E.D. Jr. Predator-prey arms races. Bioscience 49, 557-568 (1999)

12. Vermeij, G.J. Evolution in the consumer age: predators and the history of life. Paleontological Soc. Papers 8, 375-393 (2002)

13. Dietl, G.P. $\&$ Kelley, P.H. The fossil record of predator-prey arms races: coevolution and escalation hypotheses. Paleontological Soc. Papers 8, 353-374 (2002)

14. Dietl, G.P. Coevolution of a marine gastropod predators and its dangerous bivalve prey. Biol. J. Linnean Soc. 80, 409-436 (2003)

15. Hazen, R.M., Griffin, P.L., Carothers, J.M. $\&$ Szostak, J.W. Functional information and the emergence of biocomplexity. PNAS 104, 8574-8581 (2007)

16. Endler, J.A. Defense against predators. Pages 109-134. In Feder, M.E. $\&$ Lauder, G.V., eds. Predator-Prey Relationships. Chicago: The University of Chicago Press (1986)

17. Buckling, A. $\&$ Rainey, P.B. Antagonistic coevolution between a bacterium and a bacteriophage. Proc. R. Soc. Lond. B 269, 931-936 (2002)

18. Brodie, E.D. Jr., Ridenhour, B.J. $\&$ Brodie, E.D. III. The evolutionary response of predators to dangerous prey: hotspots and coldspots in the geographic mosaic of coevolution between garter snakes and newts. Evolution 56, 2067-2082 (2002)

19. Franceschi, V.R., Krokene, P., Christiansen, E. $\&$ Krekling, T. Anatomoical and chemical defenses of conifer bark against bark beetles and other pests. New Phytologist 167, 353-376 (2005)

20. Lima, S.L. Putting predators back into behavioral predator-prey interactions. TREE 17, 70-75 (2002)

21. Ofria, C. $\&$ Wilke, C.O. Avida: a software platform for research in computational evolutionary biology. Artificial Life 10, 191�229 (2004)

22. Pennock, R.T. Models, simulations, instantiations, and evidence: the case of digital evolution. J. Exp. Theor. Art. Int. 19, 29-42.

23. Lenski, R.E, Ofria, C., Pennock, R.T. $\&$ Adami, C. The evolutionary origin of complex features. Nature 423, 139-144 (2003)

24. Zaman, L., Devangam, S. $\&$ Ofria, C. Rapid host-parasite coevolution drives the production and maintenance of diversity in digital organisms. In: Krasnogor, N. $\&$ Lanzi, P.L., eds. Proceedings of the 13th Annial Conference on Genetic and Evolutionary Computation (GECCO) 219-226 (2011)

25. Fortuna, M.A., Zaman, L., Wagner. A.P. $\&$ Ofria, C. 2013. Evolving digital ecological networks. PLoS Comp. Biol. 9, e1002928 (2013)

26. Hall, A.R., Scanlan, P.D., Morgan, A.D. $\&$ Buckling, A. Host-parasite coevolutionary arms races give way to flutuating selection. Ecol. Letters 14, 635-642 (2011)

27. Gandon, S., Buckling, A., Decaestecker, T. $\&$ Day, T. Host-parasite coevolution and patterns of adaptation across time and space. arms races give way to flutuating selection. J. Evol. Biol. 21, 1861-1866 (2008)

28. Lenski, R.E. Coevolution of bacteria and phage: are the endless cycles of bacterial defenses and phage counterdefenses? J. Theor. Biol. 108, 319-325 (1984)

29. Abrams, P.A. Adaptive responses of predators to prey and prey to predators: the failure of the arms-race analogy. Evolution 40, 1229-1247 (1986)

30. Harper, E.M. Dissecting post-Palaeozoic arms races. Palaeogeogr. Palaeoclimatol. Palaeoecol 232, 322-343 (2006)

31. Johnson, M.T.J. $\&$ Stinchombe, J.R. An emerging synthesis between community ecology and evolutionary biology. TREE 22, 250-257 (2007) 

32. Lawrence, D. et al. Species interactions alter evolutionary responses to a novel environment. PLoS Biol 10(5): e1001330. doi:10.1371/journal.pbio.1001330 (2012)

33. Meyer, J.R. et al. Repeatability and contigency in the evolution of a key innovation in phage lambda. Science 335, 428-432 (2012) 

34. Bryson, D.M. $\&$ Ofira, C. Understanding Evolutionary Potential in Virtual CPU Instruction Set Architectures arXiv:1309.0719 (2013) 

35. Kamil, A.C. A synthetic approach to the study of animal intelligence. Nebraska Symposium on Motivation 7, 257-308 (1987)

\subsection{Acknowledgements} We thank D.M. Bryson and G. Wright for their assistance in developing the experimental system. This work was supported by the BEACON Center for the Study of Evolution in Action (NSF Cooperative Agreement DBI-0939454) and the Michigan State University Institute for Cyber Enabled Research. 

\subsection{Author Contributions} A.P.W conceived and conducted the study and prepared the manuscript. L.Z. helped inspire the work and devise experiments. I.D. and C.O. were involved in the study design, implementation, and analyses. All authors discussed the methods and results and commented on the manuscript.

\subsection{Author Information} Avida configuration files, datasets, and analysis scripts have been deposited in the Dryad database. Full results produced by these configuration files (approximately 20 GB) are available upon request. Reprints and permissions information is available at www.nature.com/reprints. The authors declare no competing financial interests. Correspondence and requests for materials should be addressed to A.P.W. (apwagner@msu.edu) and C.O. (ofria@msu.edu).

\clearpage

\section{Supplementary Figures}
\captionsetup[figure]{labelformat=empty}

\makeatletter
\setlength{\@fptop}{5pt}
\begin{figure}[!htbp]
\centering

\includegraphics[width=2.9in]{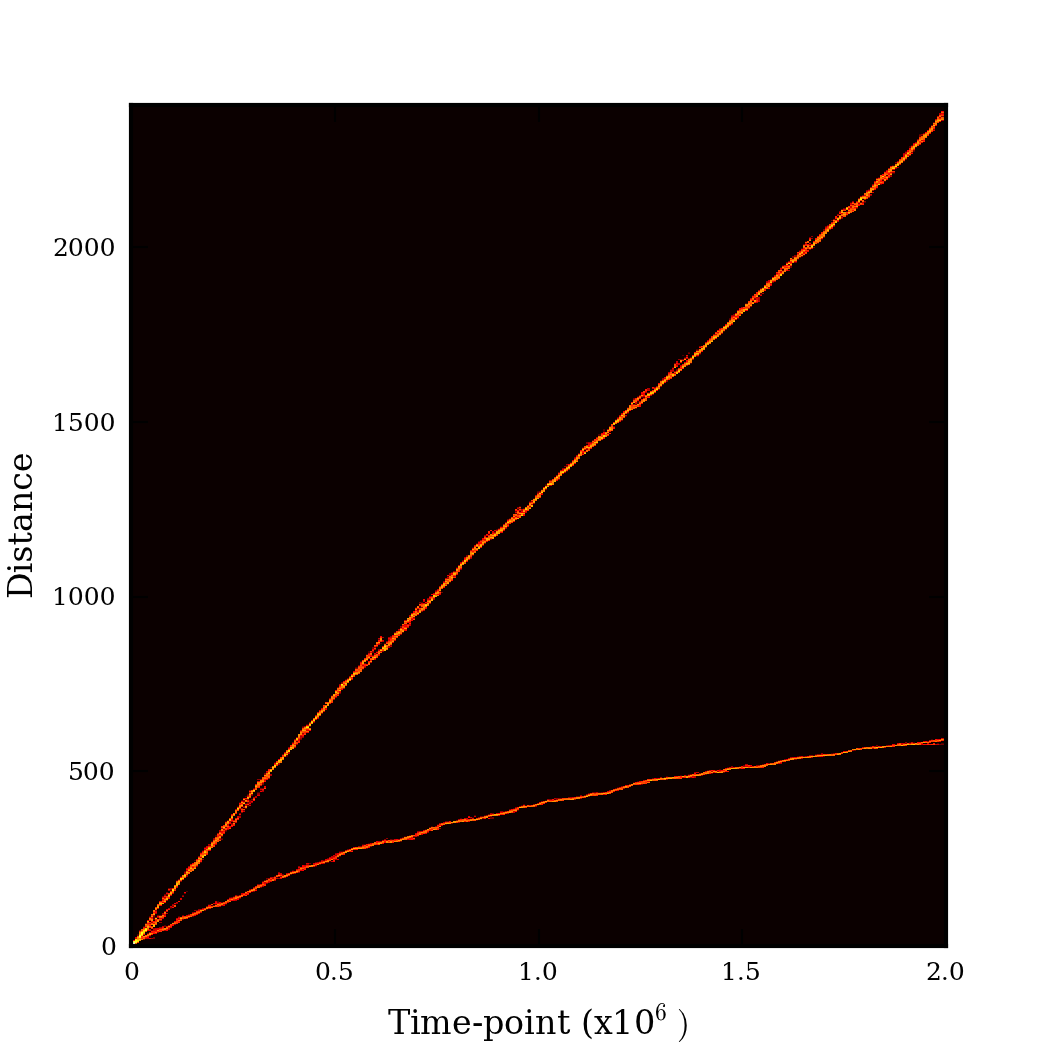}
\caption{  \textbf{Fig. S1. Predators and prey diverge genotypically as well as behaviorally.} Shown is a plot from a sample population of evolving predators and prey. X-axis indicates evolutionary time (in updates). Y-axis indicates the mutational distance for every genotype in the population to the common ancestor. Color indicates number of organisms at that depth and time. Top lineage corresponds with prey. Bottom cluster corresponds with predators. Over the course of 2 million updates of evolution, mutations created significant divergence in predators and prey genetics, as well as ‘behavioral speciation’ (e.g., as in Fig. 2 and Movie S1). Mutations tend to accumulate slower in established predator lineages because foraging inefficiencies across trophic levels slow reproductive output and generation times.
}
\label{ed1}
\end{figure}

\makeatletter
\setlength{\@fptop}{5pt}
\begin{figure}[!htbp]
\centering

\includegraphics[width=2.9in]{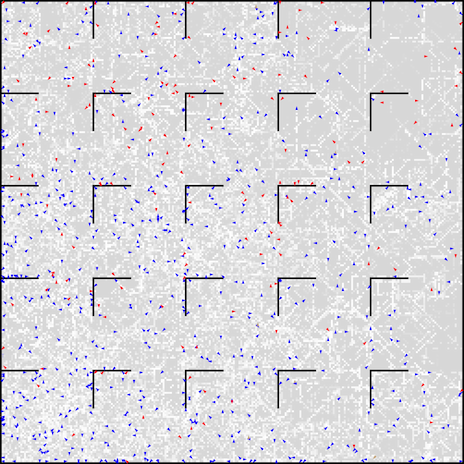}
\caption{ \textbf{Fig. S2. Predators and prey coevolved in bounded, cell-based grid-worlds.} Shown is a sample evolved population of predators (red) and prey (blue) in their 251 X 251 grid-cell environment. Black lines indicate barriers, included in some treatments (as specified in the main text), that block movement (shown here, full sized = 20 cells long on each axis, and one cell wide). Grey to white background illustrates prey forage levels by cell (grey = edible, white = consumed and regrowing). Maximum sight distance was 10 cells.
}
\label{ed2}
\end{figure}

\makeatletter
\setlength{\@fptop}{5pt}
\begin{figure}[!htbp]
\centering

\includegraphics[width=2.9in]{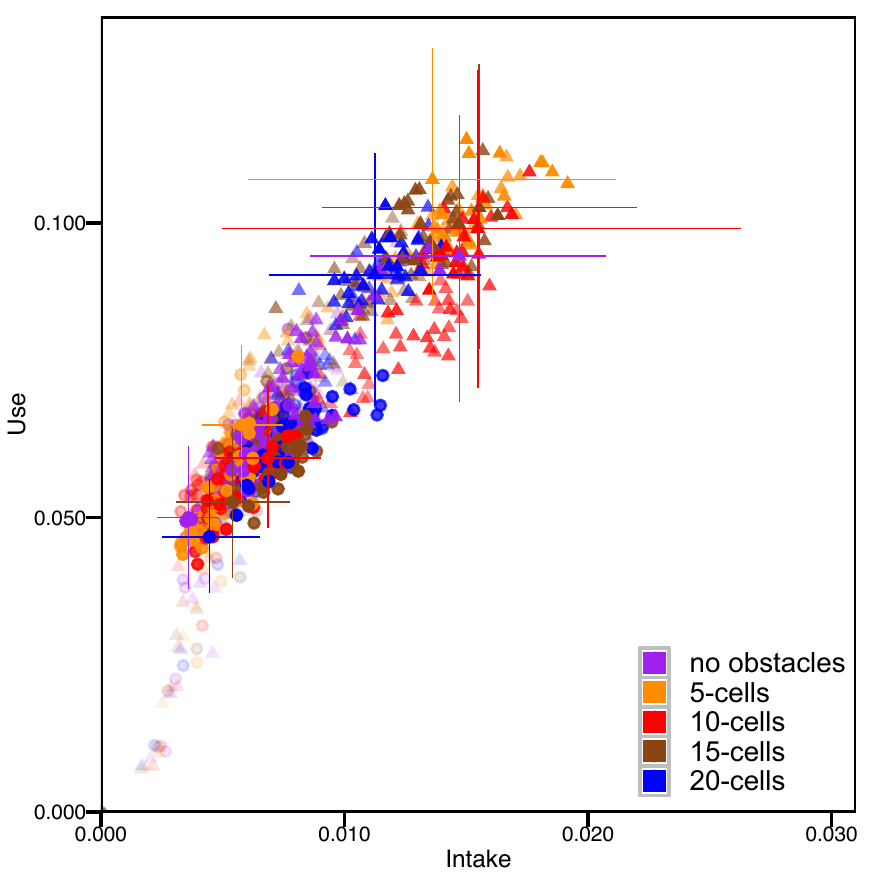}
\caption{ \textbf{Fig. S3. Behavioral intelligence and complexity did not scale with complexity of the abiotic environment.} The complexity of the abiotic environment was adjusted by adding obstacles 5, 10, 15, and 20 cells long to the environment (see Fig. S2). Sensory intake was measured as the ratio of lifetime sensor information intake to total actions taken. Behavioral intelligence was measured as the mean proportion of lifetime decisions that used sensory data about the environment. Error bars illustrate bootstrapped 95% confidence intervals around the means for the final time-point of the evolutionary trials. Points indicate mean within-treatment values at 20,000 time-point (update) intervals. Shading indicates update sampled (lighter = older). Circles indicate means for prey populations evolving without predators (none of which escaped the bottom cluster of low rates of information intake and use). Triangles represent data for prey populations coevolving with predators (all of which reached the top cluster of high complexity and intelligence). While there was no clear pattern of environmental complexity driving the evolution of prey behavioral complexity and intelligence, coevolution with predators consistently increased both measures, both within and across environmental treatments. 
}
\label{ed3}
\end{figure}

\makeatletter
\setlength{\@fptop}{5pt}
\begin{figure}[!htbp]
\centering

\includegraphics[width=2.9in]{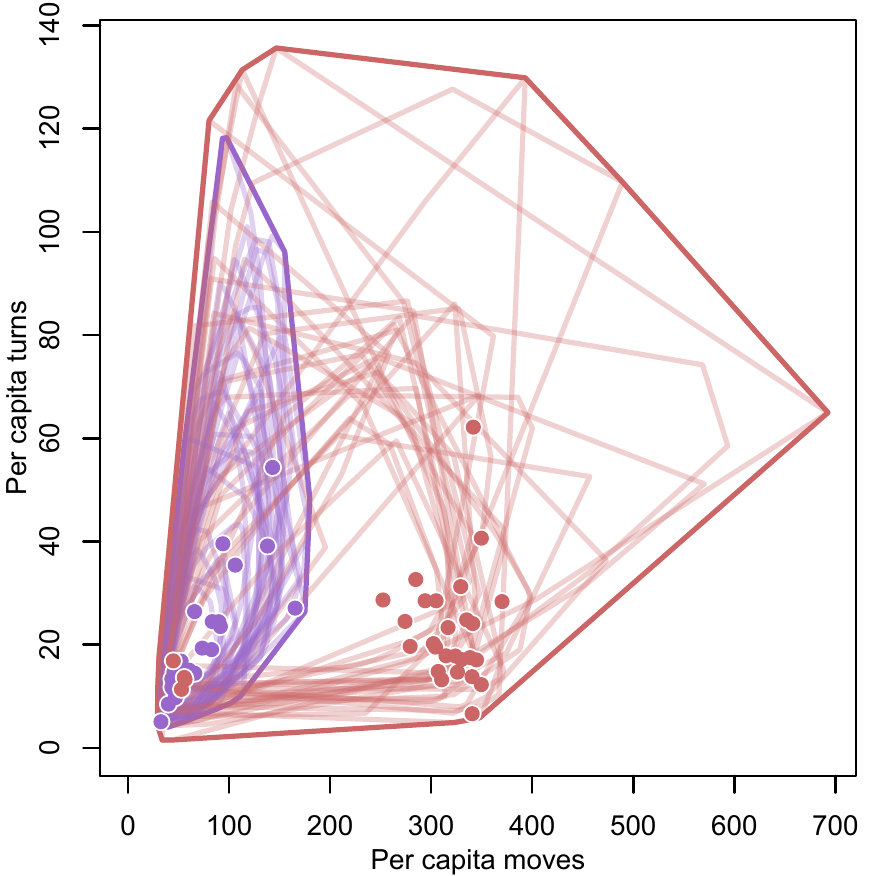}
\caption{ \textbf{Fig. S4. Prey populations coevolving with predators explore a larger area of behavioral space.} While traveling greater distances (Fig. 2), prey coevolving with predators explore larger and more varied areas of behavioral strategy space, with most settling in a an area of relatively low turns rates, but high movement rates. Populations coevolving with predators are shown in red. Prey populations evolving in the absence of predators are shown in purple. Points indicate final per-capita move-turn rates for each of the prey populations under each treatment. Dark outlines indicate cumulative convex hulls for all populations of each treatment. Lighter outlines indicate convex hulls for the areas explored by each individual population. Note that three populations coevolving with predators did not escape the low-movement behavioral space, never exploring beyond, or successfully crossing, an apparent behavioral valley bordering the area in which all naïve populations started and in which all populations evolved without predators remained (see also Fig. 2).
}
\label{ed4}
\end{figure}

\makeatletter
\setlength{\@fptop}{5pt}
\begin{figure}[!htbp]
\centering

\includegraphics[width=2.9in]{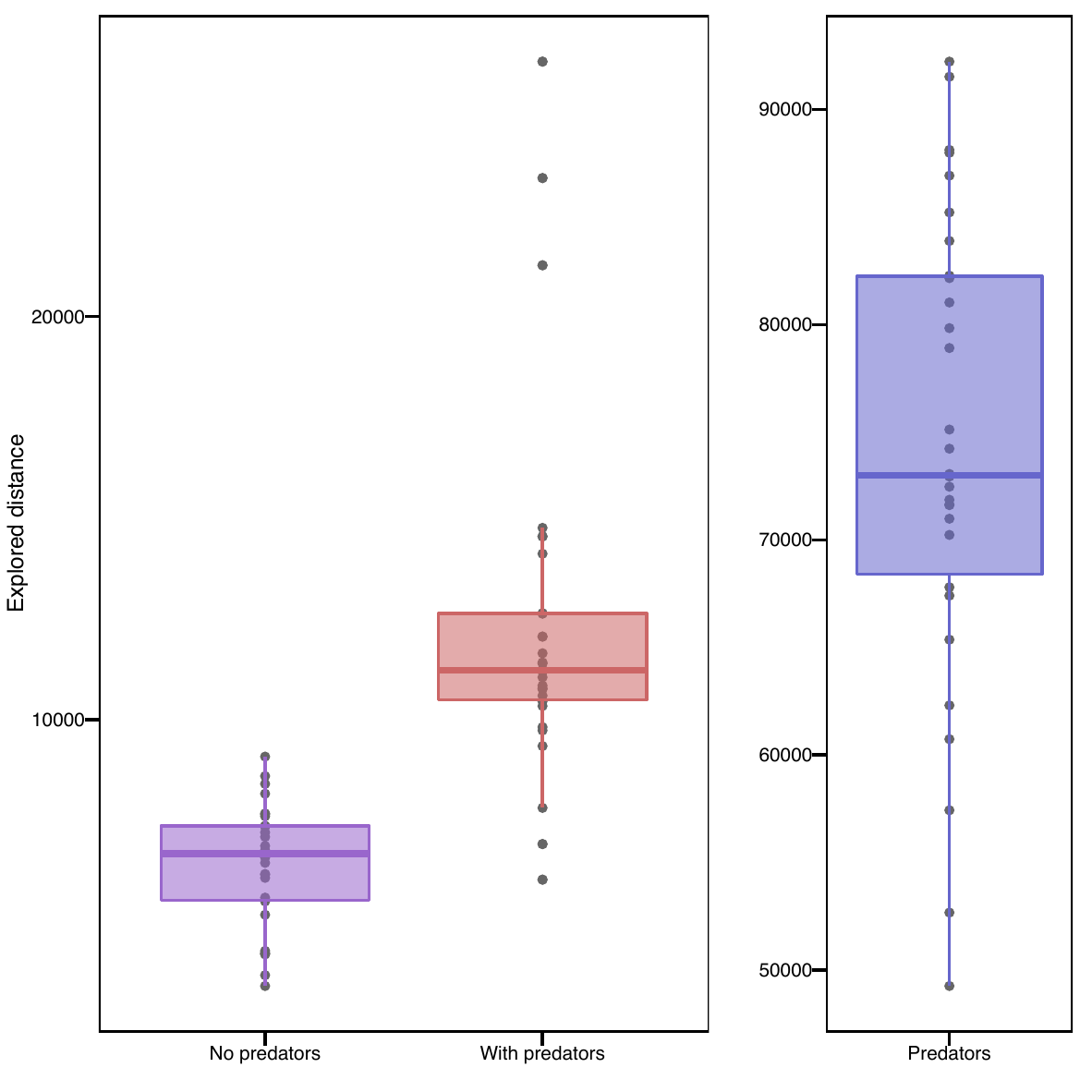}
\caption{ \textbf{Fig. S5. Predators and their prey travel farther in their explorations of behavioral space than prey populations evolving alone.} For each population, the explored distance was measured as the cumulative distance traveled over the plane defined by per-capita executions of moves and turns. Points indicate total explored distance over the full two million updates of evolution. Boxes extend from first to third quartiles. Whiskers extend from the first/third quartiles out to the highest/lowest values within one and half times the distance between the first and third quartiles. Travel distance for prey populations evolving alone and those coevolving with predators are show at left. Predator exploration of behavioral space, at a different scale, is shown on the right.
}
\label{ed5}
\end{figure}

\makeatletter
\setlength{\@fptop}{5pt}
\begin{figure}[!htbp]
\centering

\includegraphics[width=2.9in]{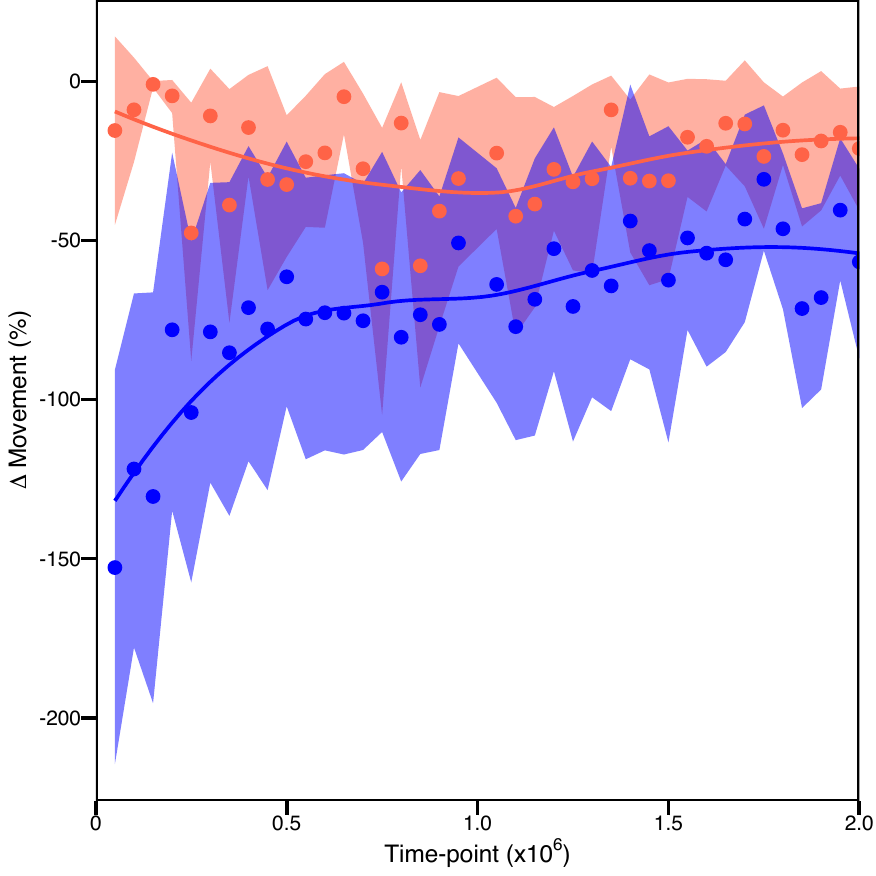}
\caption{ \textbf{Fig. S6. Prey evolve to respond to predation pressure by increasing movement, but do not increase that response over evolutionary time.} Shown (in red) are the mean per-population changes in number of steps taken by prey in environments with predators removed relative to their level of movement in environments with predators included. When predators are removed, prey consistently respond by reducing motility. However, the level of decreased movement does not change over evolutionary time. Therefore, improved anti-predator success (Fig. 2) could not have been reliant on a chase-flee movement arms race. At the same time, predators (blue) reduce levels of movement when introduced into test environments with prey removed, with the magnitude of the change stabilizing over evolutionary time. Populations were drawn from the source populations at time-point intervals of 50,000, and tested with and without the competing species removed. Data shown are from update 1000 in the test environment. Shaded regions are bootstrapped 95\% confidence intervals. Lines are LOESS fits.
}
\label{ed6}
\end{figure}

\makeatletter
\setlength{\@fptop}{5pt}
\begin{figure}[!htbp]
\centering

\includegraphics[width=2.9in]{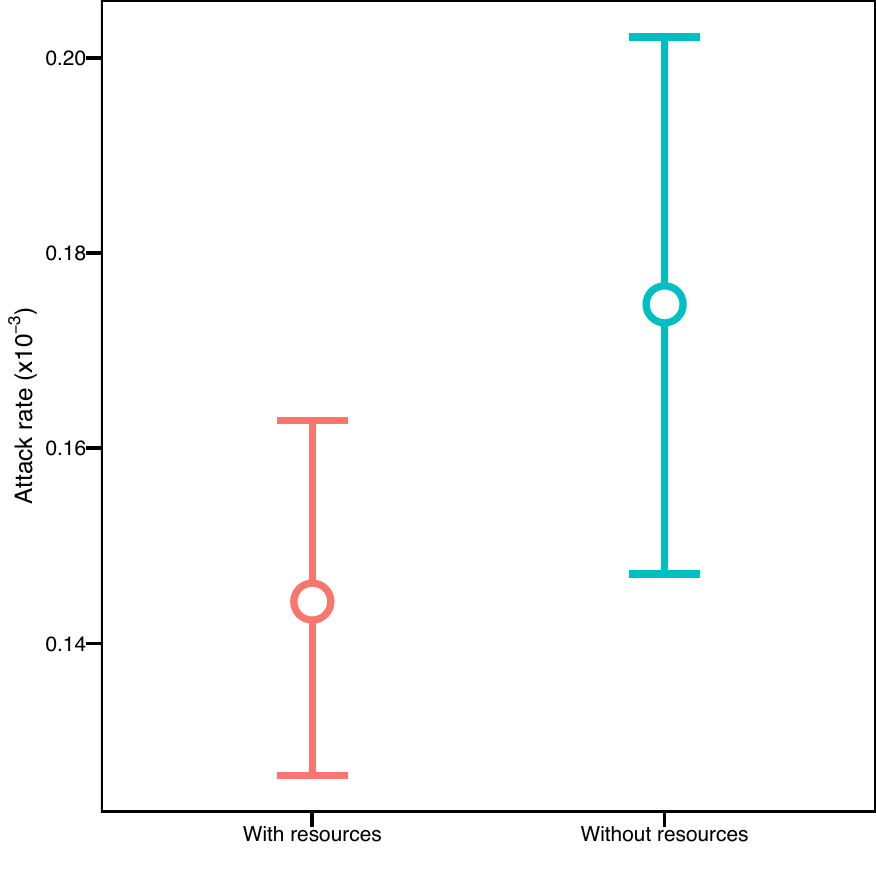}
\caption{ \textbf{Fig. S7. Prey adaptive use of sensors for finding food improves evolved predator-avoidance skills.} Prey evolved in environments without resources are less able to avoid predators, as indicated by attack rates, than those evolved in the presence of resources. After two million updates of evolution, mean predator attack rates on prey evolved in the presence of resource was 0.144 (x10\textsuperscript{-3}), 95\% CI: 0.126,0.062, compared to a mean of 0.175 (x10\textsuperscript{-3}), 95\% CI: 0.148,0.202, on prey evolved in the absence of resources. Given that prey evolved in the absence of predators are less successful in foraging than prey evolved in the presence of predators (Fig. 4), there appears to be a reciprocal evolutionary relationship in which, as prey become better at sensing and reacting to predators, they also evolve to better sense and react to resources, which further enhances evolving responsiveness to predators. Here, organisms were required to consume 10 units of resource (and prevented from consuming more). Exclusively for these treatments, for every unit collected, the metabolic rate of organisms was increased by an additive factor of one. In the resource environment, as in the main experimental treatments, prey consumed resource by feeding from cells. For the resource free environment used here, ‘resources’ were ‘consumed’ simply by moving, but no resources were removed from the environment. Thus a prey in the resource environment would increase fitness by avoiding predators while also finding and consuming resources. Contrastingly, a prey in the resource-free environment could improve its fitness simply by moving and avoiding predators. Because resources were unlimited in these two environments, prey populations were capped at 800 individuals, and the full population at 1000. Thus, in order to allow removal of spatially distributed prey food resources, this test required substantial changes to the evolutionary environment and this particular result should be viewed with some caution. Vertical bars indicate bootstrapped 95\% confidence intervals around the means (points) for the 30 trials.
}
\label{ed7}
\end{figure}

\newpage
\section{Supplementary Videos}
\captionsetup[figure]{labelformat=empty}

\makeatletter
\setlength{\@fptop}{5pt}
\begin{figure}[!htbp]
\centering

\includegraphics[width=2.9in]{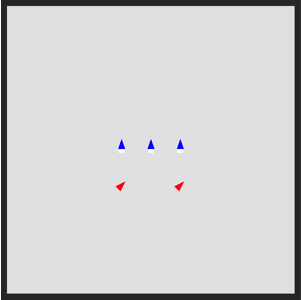}
\caption{  \textbf{Movie S1. Predators and prey evolve complex and intelligent processes for taking in, processing, and responding to information about their environment.} Shown are clones of an evolved predator (red) and prey (blue) pulled from a larger population. For this example, the predators are prevented from killing and eating the prey. While predators have evolved to look for, identify, orient toward, target, chase, and attack individual prey, prey have evolved to consume the food resources they need (grey background; white = consumed) while also avoiding predators. Neither predators nor prey can see behind them and so prey escape from predators, as in nature, is aided by frequent changes in movement directions. Sight distance is limited to 10 cells (steps), so the predators can and do lose sight of prey.
}
\label{sv1}
\end{figure}

\end{document}